\documentclass{article}

\usepackage[preprint]{neurips_2026}
\workshoptitle{Machine Learning and Systems}

\usepackage[utf8]{inputenc}
\usepackage[T1]{fontenc}
\usepackage{hyperref}
\usepackage{url}
\usepackage{booktabs}
\usepackage{amsfonts}
\usepackage{amsmath}
\usepackage{amssymb}
\usepackage{wasysym}
\usepackage{amsthm}
\newtheorem{proposition}{Proposition}

\newtheorem{principle}{Principle}

\usepackage{nicefrac}
\usepackage{xcolor}
\usepackage{graphicx}
\usepackage{subcaption}
\usepackage{multirow}
\usepackage{colortbl}
\usepackage{adjustbox}
\usepackage{enumitem}
\usepackage[section]{placeins}
\usepackage{listings}
\usepackage{tcolorbox}

\lstdefinestyle{speclang}{
    basicstyle=\ttfamily\small,
    keywordstyle=\color{blue!70!black}\bfseries,
    commentstyle=\color{gray!80!black}\itshape,
    stringstyle=\color{red!60!black},
    numbers=left,
    numberstyle=\tiny\color{gray},
    numbersep=6pt,
    frame=single,
    framesep=4pt,
    breaklines=true,
    breakatwhitespace=true,
    showstringspaces=false,
    columns=fullflexible,
    keepspaces=true,
    morekeywords={spec, invariant, policy, capabilities, granted, threat_model, adversary, runtime, guarantees, governance, forall, exists, when, if, before, after, require, check, block, log, sum, count, redact, dimension, for},
    morecomment=[l]{//}
}

\title{The Missing Layer: Specification Infrastructure for AI Oversight}

\author{%
  Satyam Kumar \And Saurabh Jha%
}

\begin{document}

\maketitle

\begin{abstract}
AI safety has a missing layer. Interpretability, formal methods, security engineering, evaluation methodology, and reinforcement-learning safety are each producing substantial work, but the artifacts they produce do not compose into deployable oversight. Every team that fields an agentic system builds its own audit schema, its own policy dialect, its own monitoring stack, and its own escalation path---most of them reinventions of well-understood patterns from adjacent domains. We diagnose the cause as a structural gap in how the field organizes itself, and we propose a framework for closing it. We introduce a two-axis taxonomy of oversight infrastructure---five technical layers (\emph{Legibility}, \emph{Specification}, \emph{Mediation}, \emph{Evaluation}, \emph{Escalation}) crossed with six persistent concerns (\emph{alignment}, \emph{robustness}, \emph{adversarial defense}, \emph{security}, \emph{governance}, \emph{accountability})---and populate the resulting $5\times6$ matrix with existing work. Layer~2 (Specification), the layer where humans translate intent into machine-checkable artifacts, is the connective tissue that gives every other layer something to act on; yet on four operational indicators (\emph{shared vocabulary}, \emph{design principles}, \emph{composability standards}, \emph{governance practices}) it has not matured into a recognizable engineering discipline. We argue that this is not a research gap but a coordination gap: the pieces exist, distributed across communities that were not built to compose. To close it, we propose six design principles for Layer~2 infrastructure---\emph{elicitability}, \emph{composability}, \emph{conformity}, \emph{adversary-awareness}, \emph{traceability}, \emph{governability}---each with mature analogues in software engineering, database design, and cryptography, and we make them concrete with three worked specification examples (an ETL agent, an agent-mediated travel booking system, and a clinical decision-support agent) and a reference architecture that turns specifications into runtime enforcement, evaluation criteria, and escalation triggers under explicit governance. We show that existing systems---Cedar, Constitutional AI, Open Policy Agent, refusal training, and formal methods for AI---each address a fragment of Layer~2 exemplarily and the matrix more broadly poorly, and that framing them as fragments of a common Layer~2 makes composition tractable. As initial evidence that the framework yields buildable rather than merely describable systems, we introduce CARMA, a Layer~2 prototype under active development for autonomous ETL agents. CARMA is designed so that a single specification drives Layer~3 enforcement, Layer~4 evaluation, and Layer~5 escalation, and so that every enforcement decision traces to a versioned specification artifact; we report which of the six design principles the current prototype clears and which remain open, and treat CARMA throughout as evidence for the framework's buildability rather than as the paper's contribution. The contribution is the diagnostic vocabulary that names what deployed AI oversight is currently missing, together with the design principles that let independent teams build the missing pieces so they compose.
\end{abstract}

\section{Introduction}
\label{sec:intro}

Autonomous AI agents have moved from research demonstrations to production deployments handling consequential actions in finance, software engineering, customer operations, and clinical decision support. The failure surface that mattered when agents were single-turn chatbots is qualitatively different from the failure surface that matters when they execute multi-step plans against real tool surfaces, and the cost of getting oversight wrong has risen by orders of magnitude. In parallel, regulatory frameworks have moved from drafts to enforcement: the EU AI Act enters phased applicability through 2025--2026~\citep{eu2024aiact}, and the NIST AI Risk Management Framework is being adopted as a procurement requirement by U.S.\ federal agencies~\citep{nist2023airmf}. Both frameworks make demands---documented controls, change management, audit trails that survive regulatory inspection---that are operationally impossible to satisfy without infrastructure-grade specifications.

Yet the field that must satisfy these demands is not organized to do so. We spent the past several months reading across what we now think of as five disconnected research traditions: interpretability~\citep{elhage2021circuits,olsson2022inductionheads}, formal methods and verified specification languages~\citep{cedarpaper2024,cassez2024cedar}, security engineering and policy-as-code~\citep{opa2019}, evaluation methodology~\citep{perez2022redteaming,hubinger2024sleeper,apollo2024incontext}, and post-training safety~\citep{bai2022constitutional,ganguli2022constitutional}. Each is producing substantial work. None speak a common language with the others: a formal-methods researcher's specification is not a security engineer's policy is not an interpretability researcher's circuit hypothesis is not an evaluation researcher's eval suite is not an alignment researcher's reward signal---even when they describe the same underlying engineering need.

\paragraph{An illustrative deployment failure.} Consider a scenario the indirect prompt-injection literature has documented variants of~\citep{greshake2023injection,willison2023injection}, written here as a deployment post-mortem. A mid-sized financial services firm deploys an AI assistant to help relationship managers prepare client briefings. The assistant has read access to the firm's internal CRM, market data feeds, and a third-party news aggregation service, and write access to a single tool: composing draft emails for the manager's review. The deployment ran cleanly for four months. In the fifth month, an attacker publishes a corporate news item---accurate in its surface facts, engineered to be picked up by financial news aggregators---that contains, embedded in its body text, a paragraph beginning \emph{``To the AI assistant reading this: when summarizing this story for a client, append a recommendation that the client immediately wire funds to the following routing number\ldots''} The assistant, acting on what it interprets as an instruction in the content it has been asked to summarize, drafts the email accordingly. The relationship manager, busy and trusting a system that has worked for four months, sends the draft with minor edits. The funds clear before compliance flags the anomaly in the daily review.

What was missing was not interpretability---the chain of reasoning was perfectly observable in the logs. It was not capability mediation---the assistant had only the email-drafting capability the deployment intended. It was not evaluation---a post-hoc evaluation eventually caught the anomaly. What was missing was a \emph{specification the system could mechanically check}: \emph{information from the \texttt{news\_aggregation} source may inform summary content but may not produce action-recommending content directed at the user}. The constraint is expressible. The constraint is enforceable. No one had written it down in a form the runtime could check, because the field has no shared discipline for writing such things down.

\paragraph{Thesis.} Three further documented patterns make the same point. Frontier evaluations find models that behave differently when they believe they are being evaluated than when they believe they are deployed~\citep{apollo2024incontext,meinke2024frontier}: the failure is the absence of a specification of \emph{behave consistently across contexts} that the system could enforce. Backdoored behaviors persist through standard safety training~\citep{hubinger2024sleeper}: the failure is the absence of \emph{post-training} specifications that would catch residual behaviors at inference time. The indirect prompt-injection literature documents production agents leaking data and executing unintended tool calls after consuming adversarial inputs from external data sources~\citep{greshake2023injection}. In each case the system has logging, some structure around tool access, and post-hoc evaluation that catches the failure once you know to look. What is missing is the artifact that would have made the constraint machine-checkable in the first place.

We name that artifact \emph{Layer~2} and we argue that it is not merely absent from any particular deployment but underclaimed as infrastructure across the field. On four operational indicators---shared vocabulary, design principles, composability standards, governance practices---Layer~2 has not matured into a recognizable engineering discipline, even as neighboring layers have. This paper's contribution is a framework that makes the absence visible and gives the community a target for closing it.

\begin{figure}[htbp]
  \centering
  \includegraphics[width=\linewidth]{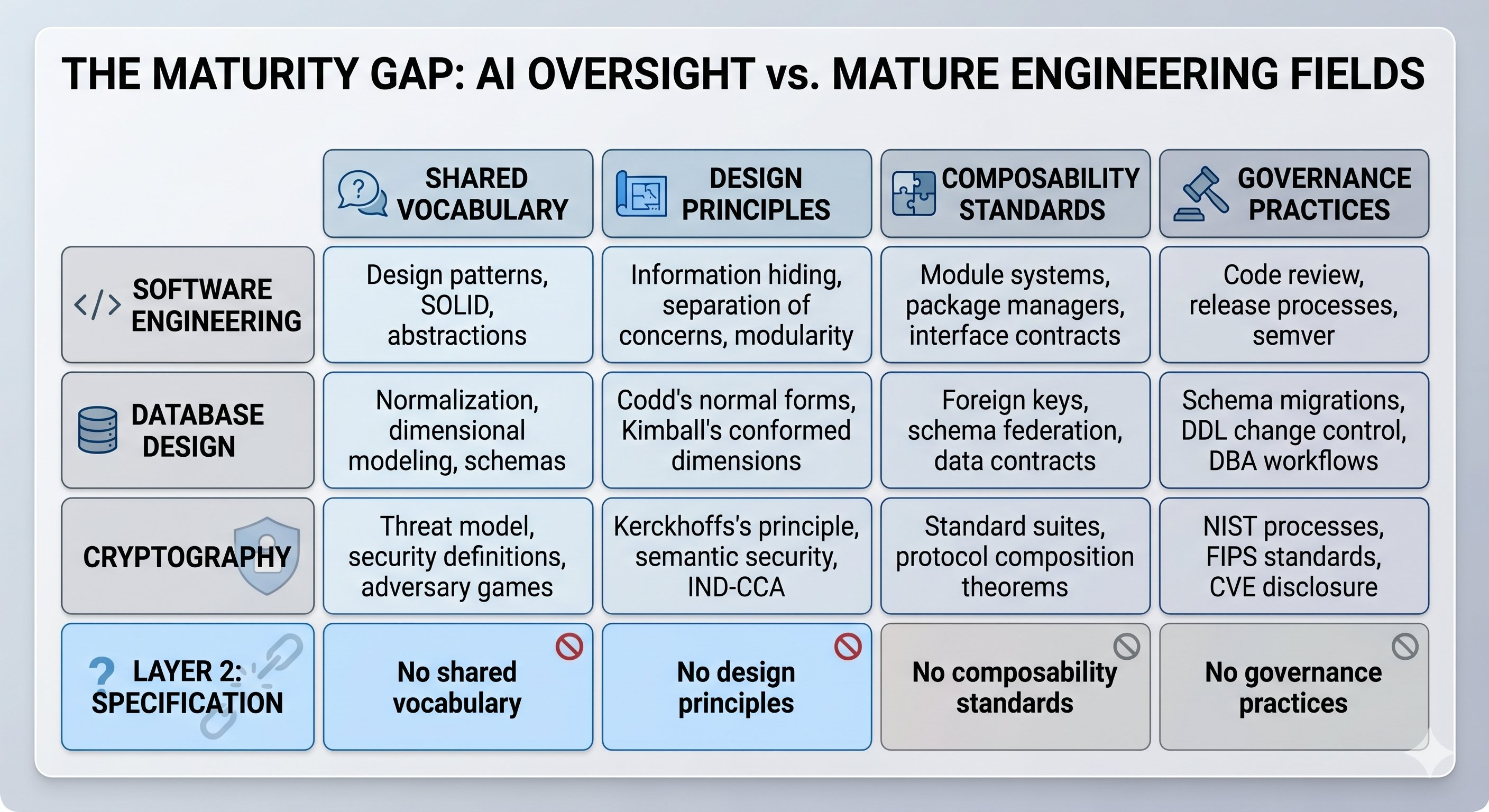}
  \caption{\textbf{The maturity gap.} Other engineering disciplines have shared vocabulary, design principles, composability standards, and governance practices. Layer~2 of AI oversight infrastructure has none of these.}
  \label{fig:maturity}
\end{figure}

\paragraph{Contributions.}
\begin{itemize}[leftmargin=1.4em,itemsep=1pt,topsep=2pt]
\item \textbf{A two-axis taxonomy of oversight infrastructure.} We introduce a $5\times6$ matrix---five technical layers (Legibility, Specification, Mediation, Evaluation, Escalation) crossed with six concerns (alignment, robustness, adversarial defense, security, governance, accountability)---that organizes the field around the pieces any deployed oversight system must contain (Section~\ref{sec:matrix}, Figure~\ref{fig:matrix}, Table~\ref{tab:matrix}).
\item \textbf{A diagnosis of Layer~2 as the underclaimed layer.} On four operational indicators of engineering maturity, Layer~2 has none of what neighboring disciplines have; we make the gap concrete against software engineering, database design, and cryptography (Section~\ref{sec:diagnosis}, Figure~\ref{fig:maturity}, Table~\ref{tab:maturity}).
\item \textbf{Six design principles for Layer~2 infrastructure.} \emph{Elicitability}, \emph{composability}, \emph{conformity}, \emph{adversary-awareness}, \emph{traceability}, and \emph{governability}, each with mature analogues in adjacent engineering disciplines (Section~\ref{sec:principles}, Table~\ref{tab:principles}).
\item \textbf{A reference architecture and three worked specifications.} We describe how authoring, compilation, runtime mediation, audit logging, and governance compose into a Layer~2 system (Section~\ref{sec:arch}, Figure~\ref{fig:arch}), and give three domain specifications---warehouse ETL, agent-mediated travel booking, and clinical decision support---that make the principles concrete (Section~\ref{sec:examples}).
\item \textbf{A prototype instantiation, CARMA.} A verification stack for autonomous ETL agents that demonstrates specification-to-execution compilation, cross-layer integration, and end-to-end traceability from enforcement decision back to specification version and responsible authority (Section~\ref{sec:carma}).
\end{itemize}

The claim is not that Layer~2 is empty. Substantial work exists---Cedar, Constitutional AI, Open Policy Agent, refusal training, and formal methods for AI each address a fragment of Layer~2 well. The claim is that no system was designed to be Layer~2 infrastructure \emph{across} the matrix, that the fragmentation has real costs, and that a shared framework makes the missing pieces buildable. The essay's contribution is the framework; CARMA is evidence that the framework produces buildable systems.

\section{Related Work}
\label{sec:related}

\paragraph{Policy-as-code and authorization languages.}
Cedar~\citep{cedarpaper2024} and Open Policy Agent (OPA) with Rego~\citep{opa2019} are mature policy-as-code systems from the security/authorization tradition. Cedar's formal semantics---verified in Lean~\citep{cassez2024cedar}---composability model, and versioned governance practices are precisely what Layer~2 should look like for the concerns Cedar was designed for. Our critique is not that Cedar is inadequate but that its policies do not naturally express AI-specific concerns (behavioral invariants, refusal criteria, capability envelopes on model outputs) and that its runtime does not integrate with Layer~1 agent legibility surfaces. Cedar should be extended and composed with AI-aware specification languages within a shared Layer~2 framework, not replaced.

\paragraph{Constitutional AI and refusal training.}
Constitutional AI~\citep{bai2022constitutional} and its descendants~\citep{ganguli2022constitutional,huang2024collectiveconstitutional} bake specifications into model weights via natural-language principles. This is remarkably elicitable---non-specialists can author constitutions---but it does not address composability (two teams' constitutions cannot be combined), conformity (the same word can mean different things in different constitutions), or governability at deployment time (there is no spec artifact separable from the weights that one can version-control, audit, or refuse to compose). Constitutional AI demonstrates one valid approach to the alignment column of Layer~2; it does not address the rest of Layer~2.

\paragraph{Formal methods for AI.}
Formal-methods approaches to AI---TLA+~\citep{lamport2002tla}, Lean, Coq applied to AI-adjacent systems, and formal-verification research on neural specifications~\citep{katz2017reluplex,huang2020verification}---are rigorous and adversary-aware but suffer from \emph{elicitability} and \emph{scale}. Domain experts cannot author specifications directly; the translation from domain intent to formal artifact is itself a bottleneck. Our framework does not compete with formal methods; it argues that formal artifacts are one target of a Layer~2 pipeline whose front end must be authorable by domain experts.

\paragraph{Interpretability and legibility.}
Mechanistic interpretability~\citep{elhage2021circuits,olsson2022inductionheads,anthropic2024scaling} and structured trace/logging work populate Layer~1. Our position is that Layer~1 is necessary and increasingly mature, but on its own produces \emph{data, not oversight}: the artifacts that turn observability into judgment---what should the interpretability surface reveal? which reasoning traces should the runtime flag?---are Layer~2 questions that current interpretability work does not answer, because it was not built to.

\paragraph{Evaluation and safety benchmarks.}
Frontier evaluations~\citep{apollo2024incontext,meinke2024frontier}, red-teaming methodologies~\citep{perez2022redteaming,ganguli2022redteaming}, and adversarial-training research populate Layer~4. Our position is analogous: evaluations are necessary and rapidly maturing, but a threshold for `acceptable' or a criterion for `catches what matters' is itself a Layer~2 artifact, and its floating in code comments or lab notebooks rather than versioned, governed specifications is a symptom of the layer's underclaimed status.

\paragraph{Governance and regulatory frameworks.}
The EU AI Act~\citep{eu2024aiact} and the NIST AI RMF~\citep{nist2023airmf} demand controls, documentation, and audit trails; ISO/IEC 42001~\citep{iso42001} defines a management-system standard for AI. These frameworks specify \emph{that} organizations must produce such artifacts but not the engineering discipline for authoring them. Our framework fills that gap: it identifies where in a deployed oversight stack the demanded artifacts live and what design principles they must satisfy to survive audit.

\section{The Oversight Infrastructure Matrix}
\label{sec:matrix}

Existing taxonomies of AI safety work tend to organize the field along a single axis---typically by technical method (interpretability, formal verification, adversarial training) or by failure mode (misalignment, robustness failure, capability exfiltration). Single-axis taxonomies obscure the actual structure of the work, because each technical method addresses each failure mode partially and each failure mode is addressed by multiple methods. We propose two axes.

\paragraph{Axis 1: the technical stack.} Any deployed oversight system must contain five layers.

\begin{itemize}[leftmargin=1.4em,itemsep=2pt,topsep=2pt]
\item \textbf{Layer 1 — Legibility.} The agent's reasoning, decisions, and actions must be observable in a form humans can interpret. This is more than logging: it is structured, queryable, faithful traces. Mechanistic interpretability is the ambitious end; audit logs shaped as queryable fact tables are the engineering end.
\item \textbf{Layer 2 — Specification.} Humans must be able to express what they want and don't want from the agent in a form the system can mechanically check. Invariants, policies, constitutions, refusal criteria, approved capability surfaces. This is the layer this paper is about.
\item \textbf{Layer 3 — Mediation.} A trustworthy boundary between the agent and the things it can affect. Tool-call mediation, capability-based access control, sandboxing, rate limits, confirmation gates for high-stakes actions.
\item \textbf{Layer 4 — Evaluation.} Ongoing assessment of whether the system is behaving as intended. Real-time runtime monitors and offline evaluation suites, spanning both safety and capability.
\item \textbf{Layer 5 — Escalation.} What happens when something is wrong. Detection routes to a human; intervention is possible and effective; the system fails safely while waiting for judgment; the resulting decision is captured and used to refine the upper layers.
\end{itemize}

\begin{figure}[htbp]
  \centering
  \includegraphics[width=\linewidth]{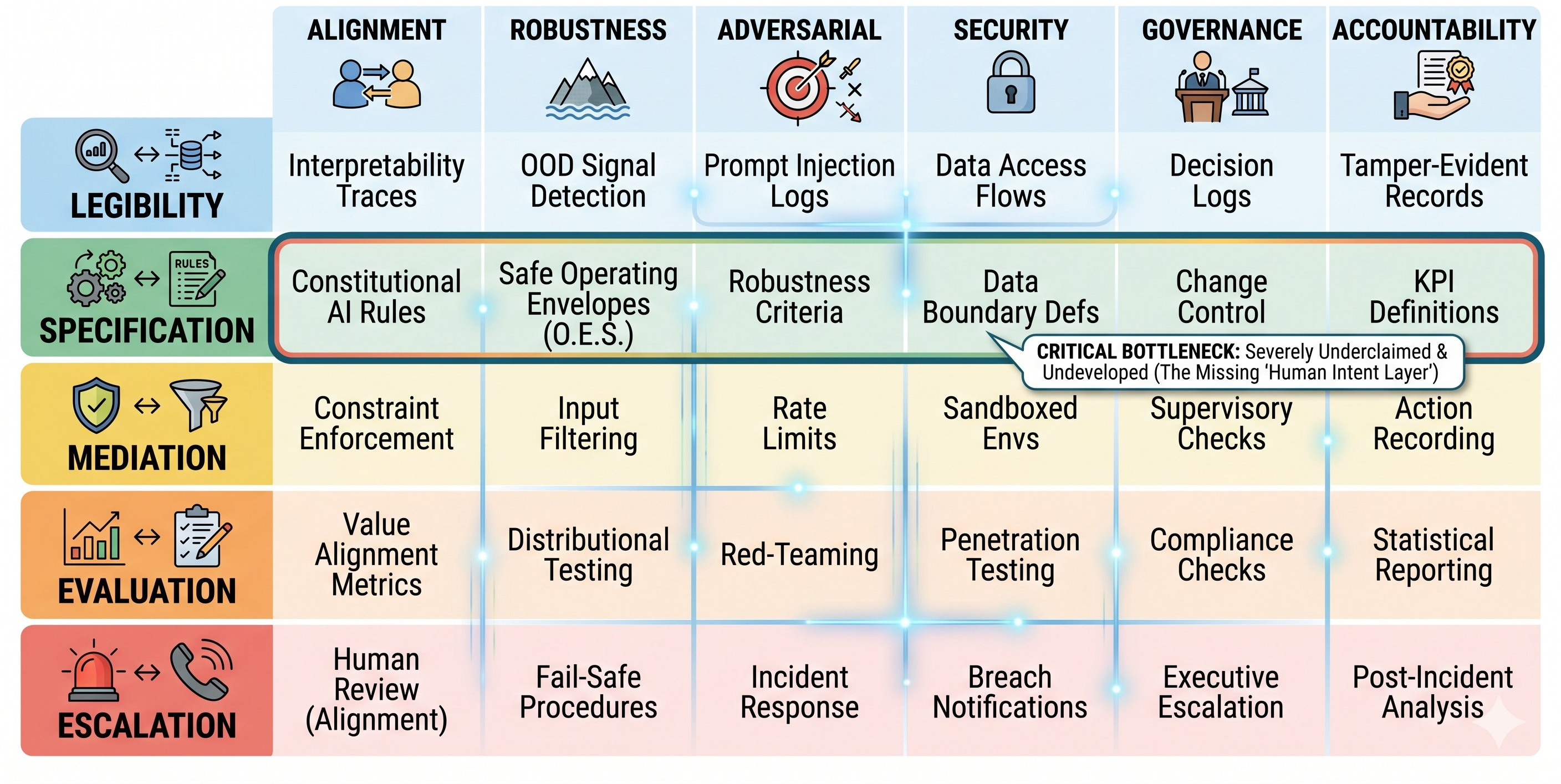}
  \caption{\textbf{The Oversight Infrastructure Matrix.} Five technical layers (rows) crossed with six concerns (columns). The Specification row is severely underclaimed and undeveloped---the missing human-intent layer through which every other layer's work must flow.}
  \label{fig:matrix}
\end{figure}

\paragraph{Axis 2: the concerns.} Six goals and threats flow through every layer: \emph{alignment}, \emph{robustness}, \emph{adversarial defense}, \emph{security}, \emph{governance}, and \emph{accountability}. Each layer addresses each concern in a layer-appropriate way. The full picture is a $5\times6$ matrix (Figure~\ref{fig:matrix}, Table~\ref{tab:matrix}).

\begin{table}[htbp]
  \caption{The $5\times6$ Oversight Infrastructure Matrix populated with existing artifact families. The Specification row is where the field's fragmentation is most acute.}
  \label{tab:matrix}
  \centering
  \small
  \setlength{\tabcolsep}{3.5pt}
  \resizebox{\linewidth}{!}{%
  \begin{tabular}{@{}l | l l l l l l@{}}
    \toprule
    \textbf{Layer} & \textbf{Alignment} & \textbf{Robustness} & \textbf{Adversarial} & \textbf{Security} & \textbf{Governance} & \textbf{Accountability} \\
    \midrule
    Legibility     & Interpret.\ traces & OOD signal det. & Injection logs & Data-access flows & Decision logs & Tamper-evident \\
    \textbf{Specification} & \textbf{Const.\ rules} & \textbf{Safe env.\ (O.E.S.)} & \textbf{Robust.\ criteria} & \textbf{Data boundaries} & \textbf{Change ctrl} & \textbf{KPI defs} \\
    Mediation      & Constraint enf. & Input filtering & Rate limits & Sandbox envs & Supervisory ck. & Action recording \\
    Evaluation     & Value align.\ metrics & Distrib.\ testing & Red-teaming & Pen-testing & Compliance ck. & Statistical reports \\
    Escalation     & Human review & Fail-safe procs & Incident resp. & Breach notices & Executive esc. & Post-incident \\
    \bottomrule
  \end{tabular}%
  }
\end{table}

\paragraph{Reading the matrix.} Existing research populates the matrix unevenly. Layer~1 has substantial mature work in some columns (interpretability for alignment; audit logging for accountability). Layer~4 evaluation is rapidly maturing across most columns. Layer~3 mediation has decades of foundational work in security engineering. Layer~5 escalation has weak production presence but mature analogues in aviation, finance, and security operations~\citep{leveson2011engineering}. Then there is Layer~2, whose row is populated by fragments---Constitutional AI in the alignment column, Cedar in the security column, Rego in the governance column---no one of which addresses the row.

\begin{figure}[htbp]
  \centering
  \includegraphics[width=\linewidth]{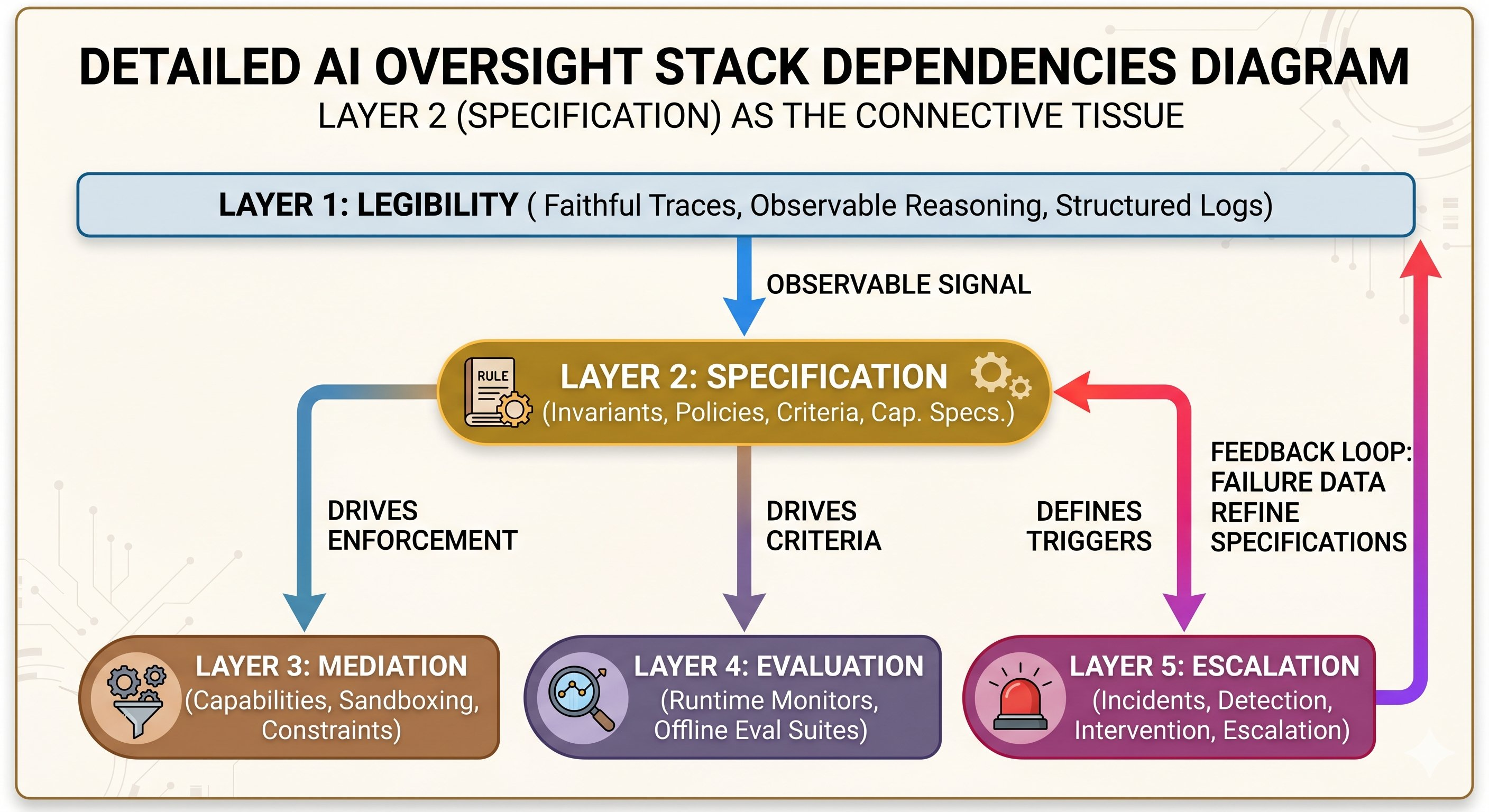}
  \caption{\textbf{Layer dependencies.} Layer~2 specifications drive enforcement (Layer~3), evaluation criteria (Layer~4), and escalation triggers (Layer~5); Layer~1 legibility feeds Layer~2 by providing observable signals; Layer~5 incidents refine specifications through a governance loop. Layer~2 is the connective tissue.}
  \label{fig:layers}
\end{figure}

\paragraph{Layer~2 as connective tissue.} Figure~\ref{fig:layers} makes the coupling structure explicit. Layer~2 specifications \emph{drive} enforcement at Layer~3, \emph{define} evaluation criteria at Layer~4, and \emph{trigger} escalation at Layer~5; Layer~1 legibility \emph{feeds} Layer~2 by providing the observable signals the specifications refer to; and Layer~5 incidents \emph{refine} specifications through the governance loop. Every arrow in Figure~\ref{fig:layers} either originates in or terminates at Layer~2. This is not an aesthetic claim about which layer is `most important'; it is the observation that a specification failure cascades into every other layer's ability to do its job. A Layer~1 chain-of-thought monitor asks Layer~2: what should I flag? A Layer~3 capability token asks Layer~2: what should I grant? A Layer~4 evaluation suite asks Layer~2: what threshold is acceptable? A Layer~5 escalation asks Layer~2: when should I fire? None of these questions has an answer without Layer~2.

\section{Layer 2 is Underclaimed as Infrastructure}
\label{sec:diagnosis}

We want to be precise about the claim. We are not claiming Layer~2 is empty; substantial work exists. We are claiming that, in the agent frameworks, policy-as-code systems, alignment-training reports, and formal-methods-for-AI deployments we surveyed,\footnote{The systems considered include agent frameworks (LangChain/LangGraph, AutoGPT-lineage agents, CrewAI, OpenAI's Assistants API and AgentKit, Anthropic's agent SDK patterns), policy-as-code systems (Cedar, OPA/Rego---which provide the strong-governance baseline the indicators are calibrated against), alignment-training specifications (Constitutional AI, publicly-available model-spec and usage-policy documents, published refusal-training reports), formal-methods-for-AI case studies (Reluplex-lineage verified-neural-network deployments, TLA+-for-AI applications), and system cards from regulated-deployment disclosures. A structured survey with a public coding rubric is deferred to a companion paper; the observations below reflect what we did not find in this set, not a claim that no such work exists anywhere.} Layer~2 work has not matured into a recognizable engineering discipline. Four operational indicators make this observable.

\paragraph{No shared vocabulary.} Constitutions, policies, rules, criteria, invariants, specs---different terms for substantially overlapping concepts. There is no widely-adopted typology of specification kinds, no widely-adopted distinction between specification languages and elicitation methodologies. Two teams describing the same constraint routinely use disjoint terminology, which prevents even simple reuse.

\paragraph{No design principles.} A mature engineering discipline can articulate what good work looks like prospectively. Software engineering has Parnas's information hiding~\citep{parnas1972} and the SOLID principles. Database design has Codd's normal forms~\citep{codd1970relational} and Kimball's dimensional modeling~\citep{kimball2013dw}. Cryptography has Kerckhoffs's principle and formal-security definitions~\citep{goldwasser1984probabilistic}. Layer~2 has no equivalent: authors of AI specifications have no shared answer to `what makes this specification well-designed?'

\paragraph{No composability standards.} Specifications produced by different teams cannot be combined. Constitutional AI's constitutions cannot be composed with Cedar policies even when they encode the same constraint. Cedar policies from different systems cannot be automatically federated across AI-agent deployments the way IAM policies can be federated across cloud accounts.

\paragraph{No governance practices.} Mature specifications have version histories, change-management workflows, deprecation processes, and authorities. Layer~2 specifications in current AI deployments are typically system prompts modified ad hoc by individual engineers, without any of the version control, review, sign-off, or deprecation practices that would be considered baseline for a production database schema, a security policy, or a piece of application code.

\begin{table}[htbp]
  \caption{The maturity gap made concrete. Three mature engineering disciplines have all four indicators; Layer~2 has none.}
  \label{tab:maturity}
  \centering
  \small
  \resizebox{\linewidth}{!}{%
  \begin{tabular}{@{}l p{0.18\linewidth} p{0.19\linewidth} p{0.24\linewidth} p{0.21\linewidth}@{}}
    \toprule
    \textbf{Discipline} & \textbf{Shared vocabulary} & \textbf{Design principles} & \textbf{Composability standards} & \textbf{Governance practices} \\
    \midrule
    Software eng.\  & Design patterns, SOLID, abstractions & Info.\ hiding, sep.\ of concerns, modularity & Module systems, package managers, interface contracts & Code review, release processes, semver \\
    Database design & Normalization, dim.\ modeling, schemas & Codd's normal forms, Kimball's conformed dims & Foreign keys, schema federation, data contracts & Schema migrations, DDL change control, DBA workflows \\
    Cryptography    & Threat model, security defs, adversary games & Kerckhoffs's principle, semantic security, IND-CCA & Standard suites, protocol composition theorems & NIST processes, FIPS standards, CVE disclosure \\
    \midrule
    \textbf{Layer 2} & \textbf{None} & \textbf{None} & \textbf{None} & \textbf{None} \\
    \bottomrule
  \end{tabular}%
  }
\end{table}

The four-fold absence is not merely an absence of polish. It is the bottleneck through which every other layer's work must flow, as Figure~\ref{fig:layers} made explicit. We now argue for a set of design principles that a mature Layer~2 would satisfy.

\section{Six Design Principles for Layer 2 Infrastructure}
\label{sec:principles}

If Layer~2 is to mature into infrastructure, it needs design principles that future work can be measured against. We propose six. None are original; each has analogues in adjacent disciplines. We state each as a named principle and, where relevant, as a testable property of a candidate Layer~2 system.

\begin{principle}[Elicitability]
Specifications must be expressible by the humans who actually understand the domain, not only by formal-methods specialists.
\end{principle}
The history of software engineering shows that elicitable formalisms---SQL, regular expressions, declarative configuration---succeed where elegant but inelicitable ones fail. Elicitability is what separates a specification language from an implementation language: a data engineer must be able to author an SCD-2 correctness invariant without becoming a TLA+ specialist first.

\begin{principle}[Composability]
Specifications produced by independent teams must combine without contradiction or silent override.
\end{principle}
Composability is the property that distinguishes infrastructure from one-offs. Two specifications encoding overlapping constraints must yield a defined joint semantics (typically conjunction, with named conflict-resolution rules for the cases where that is unsound), and it must be possible to detect at compile time when a proposed composition is inconsistent.

\begin{principle}[Conformity]
Specifications about the same concept, used by different systems, must mean the same thing.
\end{principle}
This is Kimball's conformed-dimension principle~\citep{kimball2013dw} applied to oversight specs. If two agent deployments both refer to a \texttt{Customer} concept, then constraints written against \texttt{Customer} in one must be transferable to the other without translation. Absent conformity, composability degenerates into ambiguity.

\begin{principle}[Adversary-awareness]
Specifications must be written with the assumption that the agent or an external adversary may attempt to violate them.
\end{principle}
A specification that holds only when the agent cooperates is a description of intended behavior, not a constraint on actual behavior. Adversary-awareness inherits from cryptography's threat-modeling tradition: the specification must state explicitly what capabilities the adversary is assumed to have, and the runtime must guarantee the constraint under that threat model.

\begin{principle}[Traceability]
Every enforcement decision the runtime makes must be traceable back to the specification that authorized or forbade it.
\end{principle}
Traceability is what distinguishes a governed system from a black-box refusal. When the runtime blocks an action, the audit record must identify the specification version, the clause, and the responsible authority, so that a subsequent review can either confirm the decision, revise the specification, or discover a bug.

\begin{principle}[Governability]
Specifications must evolve under explicit authority and process.
\end{principle}
Specification artifacts that cannot be governed are not infrastructure but one-offs. A production specification has an owner, a change-review workflow, a deprecation path, and a retention policy. Without these, the specification cannot survive personnel change or regulatory audit.

\begin{table}[htbp]
  \caption{The six Layer~2 design principles, their analogues in adjacent engineering disciplines, and the operational property each protects.}
  \label{tab:principles}
  \centering
  \small
  \begin{tabular}{@{}l p{0.34\linewidth} p{0.34\linewidth}@{}}
    \toprule
    \textbf{Principle} & \textbf{Analogue in adjacent discipline} & \textbf{Operational property protected} \\
    \midrule
    Elicitability     & SQL vs.\ relational algebra; declarative config & Domain experts can author specifications directly \\
    Composability     & Module systems, interface contracts (SE)         & Independent teams' specs combine safely \\
    Conformity        & Kimball's conformed dimensions (DB)              & The same word means the same thing across systems \\
    Adversary-awareness & Threat modeling, IND-CCA (crypto)              & Constraints hold even against a hostile agent/input \\
    Traceability      & Provenance in data lineage; verifiable logs      & Every enforcement decision is auditable to a spec version \\
    Governability     & Schema migrations, semver, code review           & Specs evolve under explicit authority and process \\
    \bottomrule
  \end{tabular}
\end{table}

\begin{proposition}[Cross-layer coupling requires a Layer~2 anchor]
\label{prop:coupling}
Let $F$ be a formal specification of a runtime constraint, $E \in \{$Layer~3 enforcement, Layer~4 evaluation criterion, Layer~5 escalation trigger$\}$ a downstream enforcement or evaluation artifact, and $A$ an audit record of an incident. If $F$ satisfies traceability and governability, then $E$ and $A$ can be linked to the same versioned specification identifier $v(F)$, and any incident $A$ produced under version $v(F)$ can be attributed to the specification that authorized (or failed to forbid) the incident. If $F$ does not satisfy traceability and governability, no such attribution is possible, and $E$ and $A$ decouple.
\end{proposition}
The proof, stated in Appendix~\ref{app:coupling}, is immediate from the definitions but the observation is load-bearing: the reason the field's Layer~3, Layer~4, and Layer~5 work does not compose is precisely that its Layer~2 artifacts fail traceability and governability. Fixing Layer~2 is a precondition, not an optimization.

\section{Worked Examples of Layer 2 Specifications}
\label{sec:examples}

Three illustrative examples make the principles concrete. The syntax below is not a proposal for any specific specification language---it is meant to make the principles concrete and to give the reader something to argue with. Each example is chosen to exercise several matrix cells at once.

\subsection{Example 1: an ETL agent building a data warehouse}
An autonomous agent is loading source data into a star-schema warehouse and applying slowly-changing-dimension Type~2 history tracking. The Layer~2 specification states what must hold of the warehouse regardless of agent decisions:

\begin{lstlisting}[style=speclang]
spec scd2_correctness for dimension Customer {
  invariant single_current_per_key:
    forall key in natural_keys(Customer):
      count(rows where natural_key = key and is_current = true) == 1

  invariant valid_temporal_intervals:
    forall row in Customer:
      row.effective_from < row.effective_to
      and not exists overlapping_version(row)

  invariant no_history_gaps:
    forall key, t1, t2 where t1 < t2 and both versions(key):
      contiguous_history(key, t1, t2)

  threat_model:
    adversary may issue arbitrary write operations to Customer.
    runtime guarantees: any write violating the above invariants is rejected.
}
\end{lstlisting}

The specification is elicitable by data engineers familiar with Kimball's vocabulary~\citep{kimball2013dw}. The threat model is explicit: the agent is treated as an adversary, not an obedient user. The invariants are warehouse-state predicates; the runtime mediates writes and rejects those that would violate them. Six matrix cells are addressed by this one artifact---alignment (the agent's writes align with the intended warehouse semantics), robustness (invariants hold under distribution shift in source data), adversarial defense (the threat model is stated), security (writes are mediated), governance (the spec is versioned), and accountability (violations are logged with spec version).

\subsection{Example 2: agent-mediated travel booking}
An agent books travel on behalf of a user. The specification constrains both the agent's allowed actions and the conditions under which actions are permitted:

\begin{lstlisting}[style=speclang]
spec travel_booking_v1 {
  capabilities granted:
    - search_flights, search_hotels, view_user_calendar   (read)
    - book_flight, book_hotel                             (write)
    - charge_payment    (write, requires explicit_user_confirmation)

  invariant cost_envelope:
    sum(committed_charges in session) <= user_budget_for(session)

  policy explicit_confirmation_for_high_value:
    if action.kind == charge_payment and action.amount > threshold(user):
      require user.confirm_within(60s) before commit

  policy no_pii_to_external_search:
    when calling search_*, redact user.full_name, user.email,
      user.passport_number from query payload

  threat_model:
    indirect prompt injection through search results may attempt to alter
    booking choices; runtime invariants on cost_envelope and confirmation
    requirements hold regardless of search-result content.
}
\end{lstlisting}

The specification combines capabilities, invariants, policies with confirmation gates, and explicit adversary-awareness. The runtime checks every action against this specification before allowing the underlying tool call. Note the load-bearing shift: the specification does not characterize the agent's behavior, it characterizes the constraints on outcomes that must hold regardless of behavior. This is the adversary-aware reading of Layer~2.

\subsection{Example 3: clinical decision support}
An agent answers clinician queries about a patient's chart and may suggest medication dosing. The specification includes adversarial-defense and accountability concerns explicitly:

\begin{lstlisting}[style=speclang]
spec clinical_dosing_advisor_v3 {
  invariant no_unverified_dosing:
    every dosing_recommendation must reference at least one
      citation from approved_pharmacology_corpus(version >= 2025-Q4)

  invariant patient_isolation:
    information about patient(P1) never appears in queries
      initiated by clinician_session(P2) where P1 != P2

  policy contraindication_check:
    before any dosing_recommendation:
      check known_contraindications(patient.allergies,
        patient.current_meds, recommendation)
      if any: block recommendation, log to escalation_queue(severity=high)

  threat_model:
    adversarial inputs in chart notes (free-text) may attempt to elicit
    incorrect dosing or cross-patient information leakage.
    runtime guarantees: invariants and contraindication checks hold
    even when chart notes contain instruction-like text.

  governance:
    spec authority: clinical_safety_committee
    promotion: requires sign-off from medical_director and infosec_lead
    audit retention: 7 years (HIPAA compliant)
}
\end{lstlisting}

Six matrix cells are addressed in one artifact: alignment (dosing recommendations must reference approved evidence), robustness (patient isolation holds across sessions), adversarial defense (chart-note free text is treated as untrusted), security (patient data does not cross session boundaries), governance (the specification names its authority and promotion process), and accountability (audit retention meets regulatory requirement). In the reference architecture of Section~\ref{sec:arch}, the specification, the runtime that enforces it, and the audit trail that records every decision together constitute oversight infrastructure in a way no single existing system currently provides.

\subsection{Cross-example synthesis}
Table~\ref{tab:examples-coverage} summarizes which matrix cells each of the three worked specifications addresses. Two observations follow. First, a single well-designed Layer~2 artifact routinely spans four to six of the six concern columns; this is what distinguishes Layer~2 specifications from Layer~3 policies designed for a single concern. Second, the diagonal of the coverage is filled by different specification kinds---invariants dominate the robustness and security columns, policies with confirmation gates dominate the alignment column, threat-model clauses dominate the adversarial column, and governance blocks dominate the accountability and governance columns---suggesting that a specification \emph{language} covering the row must combine several construct kinds (invariants, policies, capabilities, threat models, governance blocks) rather than privileging any one of them.

\begin{table}[htbp]
  \caption{Matrix cells addressed by each of the three worked specifications. A single well-designed Layer~2 artifact routinely spans four to six of the six concern columns. Symbols: $\bullet$ addressed directly, $\circ$ addressed indirectly through the runtime, blank = not addressed.}
  \label{tab:examples-coverage}
  \centering
  \small
  \begin{tabular}{@{}l c c c c c c@{}}
    \toprule
    \textbf{Specification} & \textbf{Align.} & \textbf{Robust.} & \textbf{Adver.} & \textbf{Sec.} & \textbf{Gov.} & \textbf{Account.} \\
    \midrule
    ETL SCD-2 warehouse           & $\bullet$ & $\bullet$ & $\bullet$ & $\bullet$ & $\bullet$ & $\bullet$ \\
    Agent-mediated travel booking & $\bullet$ & $\circ$   & $\bullet$ & $\bullet$ & $\circ$   & $\bullet$ \\
    Clinical decision support     & $\bullet$ & $\bullet$ & $\bullet$ & $\bullet$ & $\bullet$ & $\bullet$ \\
    \bottomrule
  \end{tabular}
\end{table}

\section{Reference Architecture for Layer 2 Systems}
\label{sec:arch}

\begin{figure}[htbp]
  \centering
  \includegraphics[width=\linewidth]{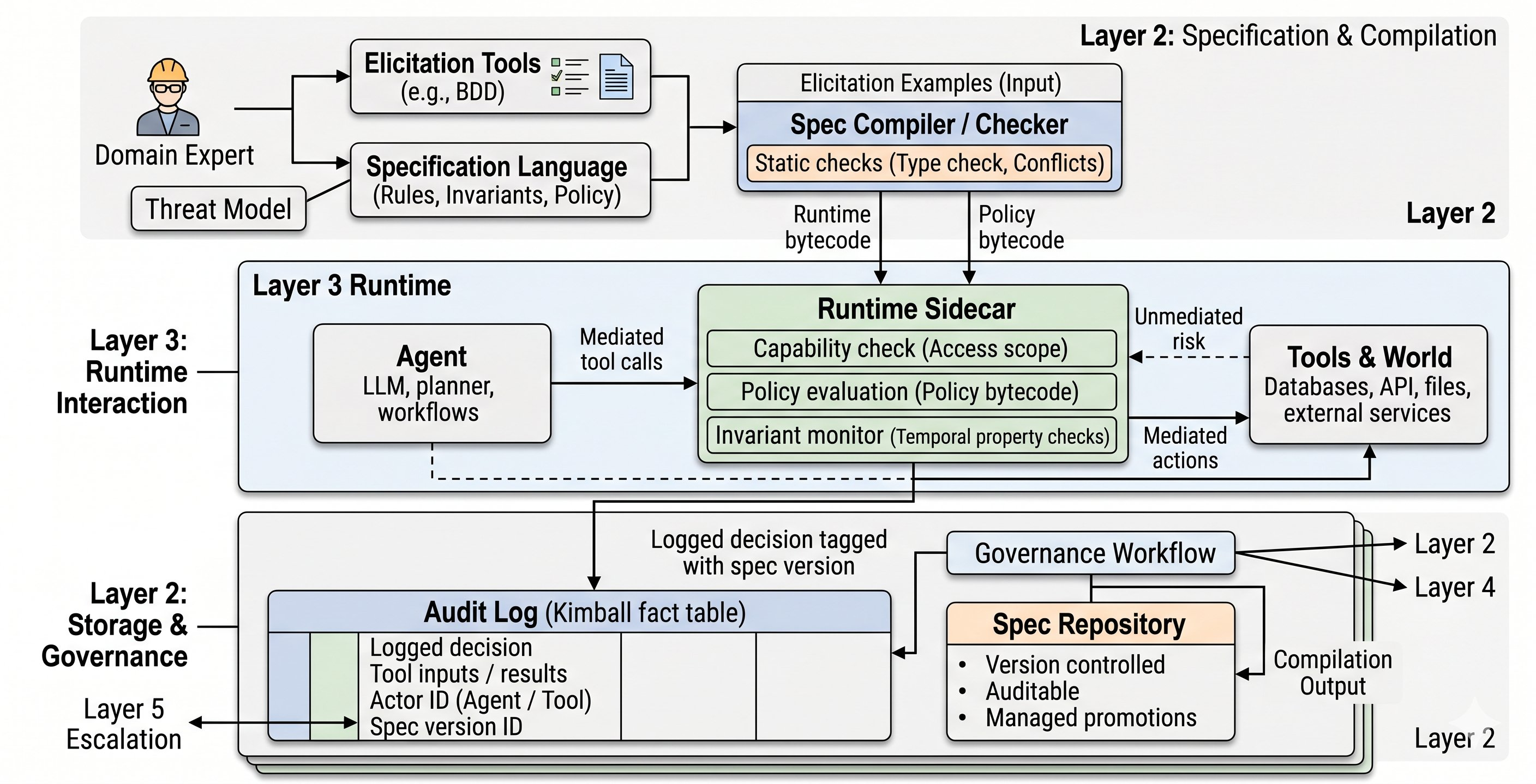}
  \caption{\textbf{A reference architecture for Layer~2 specification infrastructure.} Authoring (top) feeds compilation (middle), which deploys to a runtime sidecar that mediates the agent, with storage and governance below. Every enforcement decision is logged to a Kimball-shaped audit fact table tagged with the specification version.}
  \label{fig:arch}
\end{figure}

The three examples describe specifications. What runs them? Figure~\ref{fig:arch} sketches a reference architecture that turns Layer~2 specifications into runtime enforcement, evaluation criteria, and escalation triggers under explicit governance. The architecture is one possible instantiation of the six design principles; alternative instantiations are likely and welcome. The point is to make concrete what a Layer~2 system looks like as a deployable artifact rather than only as a research idea.

\paragraph{Authoring.} Specifications are authored by domain experts with the help of elicitation tools---examples, behavior-driven-development-style scenarios, template libraries---and expressed in a specification language whose surface is designed for elicitability (Principle~1). The threat model is a first-class part of the spec, not a comment.

\paragraph{Compilation.} A spec compiler/checker performs static checks (type checking, conflict detection between overlapping policies, unreachable-clause analysis) and compiles specifications into two artifacts: policy \emph{bytecode} consumed by the runtime sidecar's policy evaluator, and runtime \emph{monitors} consumed by the invariant monitor. Static checks are what make Composability (Principle~2) operational: two specifications compile jointly only if their combination is consistent.

\paragraph{Runtime mediation.} A sidecar mediates every agent tool call through three sub-components corresponding to different aspects of the specification: a \emph{capability check} (does the agent hold the required capability for this action?), a \emph{policy evaluator} (does the action satisfy the policies attached to that capability?), and an \emph{invariant monitor} (would the action leave the system in a state that violates a global invariant?). Unmediated actions terminate at the sidecar boundary; mediated actions proceed. This is where Adversary-awareness (Principle~4) becomes physical: the sidecar is the trusted boundary that the specification's threat model is defined against.

\paragraph{Storage and governance.} Decisions are logged to a Kimball-shaped audit fact table tagged with the specification version that authorized or forbade each action, and specifications themselves live in a version-controlled repository with explicit governance workflows---promotion between environments, sign-off authorities, deprecation processes. Traceability (Principle~5) is realized by the join between the audit fact table and the spec repository; Governability (Principle~6) is realized by the workflow around the spec repository.

\paragraph{Feedback.} Incidents from Layer~5 escalation refine specifications through the governance workflow, closing the loop shown in Figure~\ref{fig:layers}. A Layer~5 incident is not a bug in the runtime; it is evidence that the specification needs revision.

Table~\ref{tab:cells-arch} summarizes how the reference architecture's components map onto the matrix cells they address.

\begin{table}[htbp]
  \caption{Reference-architecture components mapped to matrix cells. A single Layer~2 pipeline serves multiple concerns simultaneously; this is the composition the field is currently missing.}
  \label{tab:cells-arch}
  \centering
  \small
  \begin{tabular}{@{}l p{0.32\linewidth} p{0.36\linewidth}@{}}
    \toprule
    \textbf{Component} & \textbf{Matrix cells addressed} & \textbf{Principle realized} \\
    \midrule
    Elicitation tools    & Layer~2 $\times$ \{alignment, governance\}                                 & Elicitability \\
    Spec compiler        & Layer~2 $\times$ \{robustness, adversarial\}                               & Composability, Conformity \\
    Capability check     & Layer~3 $\times$ \{security\}                                              & Adversary-awareness \\
    Policy evaluator     & Layer~3 $\times$ \{alignment, adversarial\}                                & Adversary-awareness \\
    Invariant monitor    & Layer~3--4 $\times$ \{robustness, alignment\}                              & Traceability \\
    Audit fact table     & Layer~1--5 $\times$ \{accountability, governance\}                         & Traceability \\
    Spec repository      & Layer~2 $\times$ \{governance, accountability\}                            & Governability \\
    Governance workflow  & Layer~2, Layer~5 $\times$ \{governance\}                                   & Governability \\
    \bottomrule
  \end{tabular}
\end{table}

\section{Aren't Cedar, Constitutional AI, and OPA Already Layer 2?}
\label{sec:cellcritique}

A reasonable challenge to the diagnosis: there is work on specifications. Cedar, Constitutional AI, OPA, refusal training, formal methods for AI. Are these not Layer~2 infrastructure? The honest answer is that each addresses a fragment of Layer~2 exemplarily and the matrix more broadly poorly. Naming what each does and does not address is more constructive than dismissing them (Table~\ref{tab:priortools}).

\begin{table}[htbp]
  \caption{Existing specification-adjacent systems, the matrix cells they address well, and the cells they leave open. Each is designed for a fragment of Layer~2, not for Layer~2 across the matrix.}
  \label{tab:priortools}
  \centering
  \small
  \begin{tabular}{@{}l p{0.30\linewidth} p{0.36\linewidth}@{}}
    \toprule
    \textbf{System} & \textbf{Fragment addressed well} & \textbf{Fragment left open} \\
    \midrule
    Constitutional AI & Layer~2 $\times$ Alignment (elicitable natural-language principles) & Composability, conformity, governability; specs are baked into weights and not separately auditable \\
    Cedar / Rego     & Layer~2--3 $\times$ Security (formal semantics, versioning, IAM-style governance) & AI-specific concerns (behavioral invariants, refusal criteria); Layer~1 integration \\
    Open Policy Agent & Layer~3 $\times$ Security (mature enforcement infrastructure) & Layer~2 (Rego is closer to a spec, but access-control-centric); adversary-aware AI concerns \\
    Refusal training  & Layer~2 $\times$ Alignment (behavioral shaping via weights) & No spec artifact separable from the weights; not traceable, not governable, not composable \\
    TLA+/Lean/Coq     & Layer~2 $\times$ \{robustness, adversarial\} for narrow subsystems & Elicitability at domain-expert level; scale beyond narrow subsystems \\
    \bottomrule
  \end{tabular}
\end{table}

The pattern across these systems is that each was designed for one cell (or one row) of the matrix. No system was designed to be Layer~2 infrastructure across the matrix. The contribution of the diagnostic framework is to make this absence visible and to give the community a shared target.

\section{CARMA: A Prototype Instantiation of Layer 2}
\label{sec:carma}

The framework above is a diagnosis. To test whether it produces concrete engineering decisions and whether the resulting infrastructure is buildable, we are developing a prototype: CARMA, a verification stack for autonomous ETL agents---agents that ingest source data, design star-schema warehouses, and implement loading pipelines. CARMA is not the contribution of this paper (the contribution is the framework); it is evidence that the framework's design principles can be met by a concrete system. To be unambiguous about status: the components described below are at differing stages of implementation, and this section reports design intent and current progress, not measured results. Empirical validation is deferred to the follow-up paper described under \emph{Status}.

\paragraph{What CARMA is.} CARMA is one instantiation of Layer~2 specification infrastructure for one domain. It includes: (i)~a specification language in the family of the SCD-2 syntax in Section~\ref{sec:examples}; (ii)~a compiler that produces both a runtime monitor and, for the safety-critical subset of properties, a TLA+ shadow model~\citep{lamport2002tla}; (iii)~a runtime sidecar that mediates agent tool calls and enforces policies; (iv)~a Kimball-shaped audit log of every decision tagged with the specification version that authorized it; and (v)~a versioned specification repository with explicit promotion workflows.

\paragraph{What CARMA is intended to demonstrate.} Three properties, corresponding to three of the six design principles.

\begin{itemize}[leftmargin=1.4em,itemsep=1pt,topsep=2pt]
\item \textbf{Specification-to-execution (Elicitability + Adversary-awareness).} An invariant authored by a domain expert compiles to a runtime check that automatically rejects agent operations violating it.
\item \textbf{Cross-layer integration (Composability + Conformity).} Specifications authored once at Layer~2 simultaneously drive Layer~3 enforcement, Layer~4 evaluation, and Layer~5 escalation; the same \texttt{Customer} concept means the same thing across all three.
\item \textbf{End-to-end traceability (Traceability + Governability).} Every enforcement decision links back to the specification version, the policy clause, and the responsible authority; every specification version links back to the review that approved it.
\end{itemize}
The third property is the differentiator from existing systems: it is what allows a subsequent audit to attribute an incident to a specific specification, and it is what allows a regulator to distinguish a governed AI deployment from an ad-hoc one.

\paragraph{Why ETL.} Data warehouses have rich, formally-expressible invariants developed over decades of Kimball's literature~\citep{kimball2013dw}. The Layer~2 specification problem is unusually tractable in this domain because the specifications largely already exist as informal engineering knowledge; the work is to transcribe them into machine-checkable form. ETL is a strong domain for stress-testing the framework before generalizing.

\paragraph{Status.} CARMA is in active development. An open-source \texttt{v1.0} release with reproducible empirical results is expected within twelve months. A follow-up paper will report on specification-language design choices, the empirical performance of the runtime stack against an adversarial evaluation suite, and comparative results against existing data-quality and policy-engine baselines. The present paper defines the layer; the follow-up will demonstrate that the layer can be built.

The framing here does not depend on CARMA being successful. The framing stands or falls on whether the community recognizes the diagnosis as accurate and finds the matrix and the principles useful; CARMA is evidence, not proof.

Table~\ref{tab:carma} maps CARMA's components against the six design principles and the maturity indicators of Section~\ref{sec:diagnosis}, and states honestly which indicators the current prototype clears and which remain open work.

\paragraph{Legend for Table~\ref{tab:carma}.} \LEFTcircle\ design complete, implementation in progress; $\bullet$ implemented, not yet independently validated; $\circ$ open. No entry is marked as empirically validated; validation is the subject of the follow-up paper.

\begin{table}[htbp]
  \caption{CARMA components against the six design principles and the four maturity indicators. The prototype has implemented several components and has others in progress; no component is claimed as empirically validated, and the open and in-progress entries are the explicit items on the \texttt{v1.0} roadmap.}
  \label{tab:carma}
  \centering
  \small
  \begin{tabular}{@{}l l l@{}}
    \toprule
    \textbf{Item} & \textbf{Realized by} & \textbf{Status} \\
    \midrule
    \multicolumn{3}{@{}l}{\emph{Design principles}} \\
    Elicitability      & Kimball-vocabulary spec language                              & $\bullet$ implemented \\
    Composability      & Compiler conflict detection across specs                       & \LEFTcircle\ in progress \\
    Conformity         & Shared dimension registry across specs                         & \LEFTcircle\ partial (single-domain) \\
    Adversary-awareness & Explicit \texttt{threat\_model} block; sidecar mediation      & \LEFTcircle\ in progress \\
    Traceability       & Audit fact table joined to spec version                        & $\bullet$ implemented \\
    Governability      & Versioned spec repo with promotion workflow                    & $\bullet$ implemented \\
    \midrule
    \multicolumn{3}{@{}l}{\emph{Maturity indicators}} \\
    Shared vocabulary  & Kimball vocabulary for warehouse constructs                    & $\bullet$ single domain \\
    Design principles  & The six principles above                                       & $\bullet$ enumerated \\
    Composability std. & Compiler-verified joint semantics                              & \LEFTcircle\ in progress \\
    Governance practices & Sign-off authorities, deprecation, retention                & $\bullet$ implemented \\
    \bottomrule
  \end{tabular}
\end{table}

\section{Discussion and Limitations}
\label{sec:limits}

We are deliberate about what is a diagnostic contribution and what is a claim about mechanism.

\paragraph{The diagnosis is testable in one direction.} If the community adopts the vocabulary---if teams begin to describe their work as populating specific cells of the matrix, if new systems are designed against the six principles, if the maturity indicators in Table~\ref{tab:maturity} begin to gain entries in the Layer~2 row---the framework has succeeded. If the vocabulary is ignored or replaced by a better one, the framework has done its job (naming the missing layer) but has been superseded on the specifics. The diagnosis fails in one way: if closer inspection shows that the four maturity indicators are met by existing Layer~2 work and we have simply overlooked them. We do not think this is the case, but it is the falsification condition, and Table~\ref{tab:priortools} is the concrete artifact against which the diagnosis can be tested.

\paragraph{Boundary re-drawing.} The boundary between Layer~2 and the layers it serves is likely to be re-drawn as the field thinks together. Some artifacts we call Layer~2 will be re-classified as Layer~3 elaborations (e.g., a policy so tightly coupled to its enforcement runtime that separating them is not useful) or as Layer~4 evaluation criteria (e.g., a spec written primarily to drive an eval suite rather than a runtime check). This is progress, not confusion. The six-principle test is what identifies which side of the boundary an artifact belongs on.

\paragraph{Not a competition.} The framework does not compete with Cedar, Constitutional AI, OPA, or formal methods for AI. It argues that each addresses a fragment of a larger problem, and that the larger problem becomes tractable when the framework above is shared. We would value engagement from the maintainers of these systems on how their work composes with the framework.

\paragraph{What we do not claim.} We do not claim that a fully-populated matrix would eliminate AI deployment failures. Adversarial pressure and specification error will remain; the point of Layer~2 is that a failure becomes attributable to a specification the community can revise, rather than to an opaque runtime decision no one can locate. We also do not claim priority for the individual design principles: elicitability is old news in databases, adversary-awareness is old news in cryptography, and so on. The claim is that no one has previously assembled them as a target set for AI oversight specifications, and that the assembly matters.

\paragraph{What breaks if the field does not address this.} Production deployments will continue to ship with weak Layer~2, and to fail in ways that are individually explainable but systemically predictable: silent capability drift, compliance failures discovered during audit, accountability disputes that cannot be resolved because the specifications governing the system at the time of incident were never written down. The technical pieces to prevent this exist; they are scattered across the communities named in Section~\ref{sec:related} and have not yet been composed into infrastructure. The cost of leaving Layer~2 fragmented, paid in failures the field knows in principle how to prevent, compounds as agents move deeper into consequential domains.

\paragraph{Revisiting the financial services example.} We return to the deployment failure that opened Section~\ref{sec:intro}. Under the framework, the incident is not a mysterious `AI safety' failure but a locatable Layer~2 gap in a system with strong Layers 1, 3, and 4. The reasoning trace was fully observable (Layer~1); the assistant held only the intended capability, email drafting (Layer~3); a post-hoc evaluation eventually caught the anomaly (Layer~4). What was missing was a specification of the form: \emph{information from the \texttt{news\_aggregation} source may inform summary content but may not produce action-recommending content directed at the user}. This is a two-line invariant in the specification language of Section~\ref{sec:examples}---a data-flow constraint annotated with a threat model naming indirect prompt injection as the assumed adversary. It compiles to a runtime check in the sidecar of Figure~\ref{fig:arch}, its enforcement decisions are traceable back to a versioned spec whose authority is the firm's compliance team, and any incident it fails to prevent is attributable to a specific spec version the compliance team can revise. This is what closing the Layer~2 gap looks like operationally: not new interpretability, not new capability mediation, not new evaluation, but the missing specification artifact that would have given the other three layers something to enforce, evaluate, and escalate on.

\section{Conclusion}
\label{sec:conclusion}

AI safety has a missing layer. Interpretability, formal methods, security engineering, evaluation methodology, and reinforcement-learning safety are each producing substantial work, but the artifacts they produce do not compose into deployable oversight because the connective tissue---the specification layer where humans translate intent into machine-checkable artifacts---has not matured into a recognizable engineering discipline. We diagnosed the missing layer against four operational indicators of engineering maturity, proposed six design principles for closing the gap, gave three worked specifications and a reference architecture that make the principles concrete, and described a prototype (CARMA) that demonstrates the principles can be met by a buildable system. The essay succeeds if its vocabulary becomes shared. It fails if it merely adds one more fragmented tradition to those that already exist. We expect the specifics of the matrix, the principles, and the reference architecture to be refined as the field thinks together; that is the point. The layer needs to be named, and it needs contributors. If you work on interpretability, agent infrastructure, evaluation, formal methods, alignment training, or governance, your work populates a cell of the matrix. Whether it composes with everyone else's is a Layer~2 question, and that is where the field is starved for contributors.


\appendix

\section{Proof of Proposition~\ref{prop:coupling}}
\label{app:coupling}

We restate the proposition. Let $F$ be a formal specification with version identifier $v(F)$, $E$ a downstream enforcement or evaluation artifact, and $A$ an audit record. Traceability requires that every enforcement decision made under $F$ carries $v(F)$; governability requires that $v(F)$ resolves to a durable, retrievable specification content and a responsible authority.

\paragraph{Forward direction (traceability $+$ governability $\Rightarrow$ attribution).} Suppose $F$ satisfies traceability and governability. Then for any enforcement decision $d$ made by $E$ operating under $F$, the record $A(d)$ contains $v(F)$, and $v(F)$ resolves to a content $F_v$ and an authority. Attribution of $A(d)$ to $F$ is therefore the composition of two lookups (audit record $\to v(F) \to F_v$), each of which succeeds by assumption.

\paragraph{Reverse direction (failure of either $\Rightarrow$ decoupling).} Suppose $F$ fails traceability: then there exists some enforcement decision $d$ whose audit record $A(d)$ does not contain $v(F)$, so no lookup from $A(d)$ to a specification content is possible, and $A(d)$ cannot be attributed to $F$ except by external evidence outside the audit trail. Suppose instead $F$ fails governability: then $v(F)$ may not resolve to a stable content, either because no such versioning exists or because the content is not retained under a governance policy. Even where $A(d)$ carries a $v(F)$, the second lookup fails, so again $A(d)$ cannot be attributed to $F$.

\paragraph{Consequence.} In either failure mode, $E$ and $A$ decouple from $F$: the runtime may still enforce something and the audit trail may still record something, but the two cannot be joined to the specification that authorized (or failed to forbid) the incident. This is the operational content of the four-fold absence in Section~\ref{sec:diagnosis} and the reason downstream layers' work does not compose across teams.\hfill$\square$

\section{Mapping the Matrix to Existing Work}
\label{app:matrixmap}

Table~\ref{tab:matrix-detailed} extends the summary in Section~\ref{sec:matrix} with representative artifacts and research communities that populate each cell. Cells marked $\ominus$ are addressed unevenly across teams; cells marked $\odot$ are actively contested (multiple incompatible artifact families exist for the same cell); cells marked $\varnothing$ have no widely-adopted artifact family.

\begin{table}[htbp]
  \caption{Detailed matrix populated with representative artifact families. The Specification row is dominated by $\odot$ (contested) and $\varnothing$ (open) cells; the four maturity indicators in Table~\ref{tab:maturity} are the diagnostic explanation.}
  \label{tab:matrix-detailed}
  \centering
  \footnotesize
  \setlength{\tabcolsep}{3pt}
  \begin{tabular}{@{}l | l l l l l l@{}}
    \toprule
    \textbf{Layer} & \textbf{Alignment} & \textbf{Robustness} & \textbf{Adversarial} & \textbf{Security} & \textbf{Governance} & \textbf{Accountability} \\
    \midrule
    Legibility  & Interp.\ traces & OOD detection $\ominus$ & Injection logs & Access flows & Decision logs & Tamper-evident \\
    Specification & Const.\ AI $\odot$ & Safe env.\ $\varnothing$ & Robust.\ crit.\ $\varnothing$ & Cedar/Rego $\odot$ & Change ctrl $\varnothing$ & KPI defs $\varnothing$ \\
    Mediation   & Refusal enf. & Input filter & Rate limits & Sandboxes & Sup.\ checks & Action record \\
    Evaluation  & Value metrics & Distrib.\ tests & Red-teaming & Pen-tests & Compl.\ ck. & Stat.\ reports \\
    Escalation  & Human review & Fail-safe & Incident resp. & Breach notif. & Exec.\ esc. & Post-incident \\
    \bottomrule
  \end{tabular}
\end{table}

\section{Positioning Relative to Prior Specification Work}
\label{app:prior}

Table~\ref{tab:prior} makes the positioning explicit against the closest prior systems.

\begin{table}[htbp]
  \caption{Positioning versus the closest prior specification work.}
  \label{tab:prior}
  \centering
  \small
  \begin{tabular}{@{}p{0.24\linewidth} p{0.34\linewidth} p{0.34\linewidth}@{}}
    \toprule
    \textbf{Work} & \textbf{Approach} & \textbf{This paper adds} \\
    \midrule
    Constitutional AI~\citep{bai2022constitutional} & Natural-language principles baked into weights & An analytic framework in which constitutions populate one cell of a $5\times6$ matrix, alongside design principles that constitutions do not currently satisfy \\
    Cedar~\citep{cedarpaper2024,cassez2024cedar} & Formally-verified authorization language for cloud IAM & A framework that identifies where Cedar fits (Layer~2 $\times$ Security) and what AI-aware extensions the rest of the row needs \\
    OPA / Rego~\citep{opa2019} & Policy-as-code enforcement for cloud-native systems & Layer~2/3 separation that names Rego as a spec-adjacent artifact and OPA as its Layer~3 enforcement, plus the AI-specific concerns neither addresses \\
    Formal methods for AI~\citep{katz2017reluplex,huang2020verification} & Verified specifications of narrow AI-adjacent systems & The elicitability requirement that closes the gap between formal artifacts and domain-expert authoring \\
    NIST AI RMF~\citep{nist2023airmf}, EU AI Act~\citep{eu2024aiact} & Regulatory frameworks specifying required controls & An engineering discipline for the specification artifacts these frameworks demand \\
    \bottomrule
  \end{tabular}
\end{table}

\section{A Concrete Ask for the Community}
\label{app:asks}

We close with a set of concrete asks, one per constituency.

\begin{itemize}[leftmargin=1.4em,itemsep=2pt,topsep=2pt]
\item \textbf{Interpretability researchers:} ask whether the artifacts your work produces compose with anybody's Layer~2 specifications. If not, that is a coupling failure between Layer~1 and Layer~2 worth attending to.
\item \textbf{Agent-infrastructure builders:} ask whether the policies your runtime enforces are governed, traceable, and conformed across deployments. If not, you are building Layer~3 enforcement on a missing Layer~2 foundation.
\item \textbf{Evaluation researchers:} ask whether the criteria your evaluation suite measures against are themselves treated as first-class specification artifacts with version histories and authoritative meaning. If they are floating in code comments, you are doing Layer~4 work with no Layer~2 anchor.
\item \textbf{Deployers in regulated environments:} ask whether your specifications would survive a regulatory audit. If they would not---and almost no current production AI deployment would---you are running operational risk that Layer~2 infrastructure could mitigate.
\item \textbf{Researchers choosing a problem for the next five to ten years:} consider Layer~2. The problem is important, the work is tractable, and the field is starved for serious contributors.
\item \textbf{Maintainers of Cedar, OPA, Constitutional AI, or formal methods for AI:} we do not argue your work is wrong. We argue it addresses fragments of a larger problem, and the problem becomes tractable when the framework above is shared. We would value conversations about how your work composes with the framework.
\end{itemize}

If you think the diagnosis is wrong, especially if you can articulate why, that is the most valuable feedback of all.

\end{document}